\documentclass[twocolumn,showpacs,showkeys,amsmath,amssymb,floatfix]{revtex4}

\usepackage{graphicx}
\usepackage{dcolumn}

\begin{document}

\title{Electronic states and self-doping at a 45$^{\circ}$ YBa$_2$Cu$_3$O$_7$
grain boundary}

\author{U.~Schwingenschl\"ogl$^{1,2}$ and C.~Schuster$^1$}
\affiliation{$^1$Institut f\"ur Physik, Universit\"at Augsburg, 86135 Augsburg, Germany,} 
\affiliation{$^2$KAUST, PCSE Division, P.O. Box 55455, Jeddah 21534, Saudi Arabia}
\date{\today}

\begin{abstract}
The charge redistribution at grain boundaries determines the 
applicability of high-T$_c$ superconductors in electronic devices,
because the transport across the grains can be hindered considerably.
We investigate the local charge transfer and the modification of the electronic
states in the vicinity of the grain-grain interface by first principles
calculations for a (normal-state) 45$^{\circ}$-tilted [001] grain
boundary in YBa$_2$Cu$_3$O$_7$. Our results explain the suppressed interface
transport and the influence of grain boundary doping in a {\it quantitative} manner,
in accordance with the experimental situation. The charge redistribution is found
to be strongly inhomogeneous, which has a substantial effect on transport properties
since it gives rise to a self-doping of $0.10\pm0.02$ holes per Cu atom.
\end{abstract}


\maketitle


Electronic transport in wires and tapes from high-T$_c$ compounds is seriously suppressed
by structural defects and interfaces. In particular, grain boundaries have been identified
as the main limiting factor, which determines the critical current in bulk samples \cite{dim88}.
Grain boundaries in YBa$_2$Cu$_3$O$_{7-\delta}$ (YBCO) seem to be depleted of charge
carriers \cite{browning93,babcock94}. For this reason, the supercurrent density can be
enhanced by locally overdoping the superconductor (by Ca substitution) close to grain
boundaries while keeping the grains optimally doped \cite{schmehl99}. The overdoping
prevents a reduction of the critical temperature \cite{hammerl00}, which indicates that
suppression of the critical current at grain boundaries is connected to interface charging
and/or distortions of the local electronic band structure. The latter would also explain
that a perovskite superconductor behaves different from conventional or intermetallic
superconductors, like MgB$_2$, where the supercurrent density is hardly affected \cite{samanta02}.
Yet, there is evidence for Josephson junction type weak link grain
boundaries in MgB$_2$ \cite{khare05}.

In perovskite superconductors, the critical current can drop down to 1\% of the bulk
value for a misorientation angle of 10$^{\circ}$ \cite{dim90,ivanov91,amrein95}.
For technological applications, optimization of low angle grain boundaries is mandatory
and, therefore, addressed by various groups \cite{vere00,eve04,weber06}. Large angle grain
boundaries, on the other hand, are of practical interest as they form excellent Josephson
junctions. Using bicrystal substrates or the biepitaxial principle, based on two different
substrates, 45$^\circ$ tilted [001] grain boundaries can be grown with a high accuracy
\cite{hil02,lam08}. Because substrate effects can be eliminated from transport measurements
\cite{ran04}, such Josephson junctions are widely used to gain insight into the superconducting
state (e.g., the properties and symmetry of the gap function, s-wave vs.\ d-wave)
\cite{tagl07,lam08}. In addition, mesoscopic fluctuations of the magnetoconductance point
to a coexistence of supercurrent and quasiparticle current \cite{tagl09}.

It is surprising that a first principles electronic structure calculation,
which can account for structural details at grain boundaries,
is missing so far. Probably, this fact is connected to a tremendous demand on CPU time. For
the simpler Ti-based perovskite insulators SrTiO$_3$ and BaTiO$_3$, two common substrates
for YBCO, analogous studies of grain boundaries have been performed by Klie
{\it et al.} \cite{klie03}, using the Thomas--Fermi screening approach, and by Imaeda
{\it et al.} \cite{ima08}, using density functional theory (DFT). Moreover,
superconductor-metal interfaces, which are closely related to grain boundaries in high-T$_c$
materials \cite{emig97,nikolic02}, have been addressed by DFT calculations. For YBCO-Pd
interfaces, as a prototypical example, a net charge transfer of $\sim0.1$ holes
per Cu atom off the CuO$_2$ planes has been established \cite{US_epl}, showing only a weak
dependance on the interface orientation \cite{US_apl}.
\begin{figure}[b]
\includegraphics[width=0.45\textwidth]{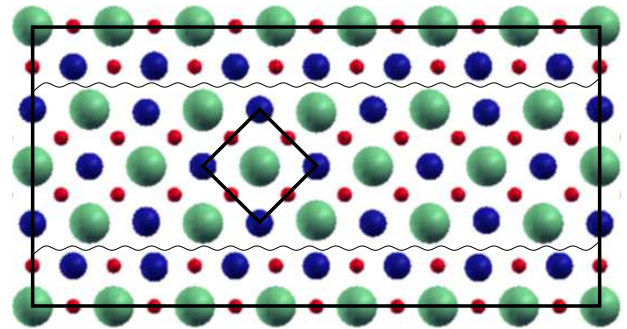}
\caption{(Color online) Structural setup used for simulating a 45$^{\circ}$ YBCO
grain boundary (wiggly line), in a projection along the YBCO $c$-axis. Large mint-green
spheres represent Y/Ba, medium blue spheres Cu, and small red spheres
O. Whereas the square highlights an YBCO unit cell, the
outer rectangle indicates the boundaries of the supercell.}
\label{fig1}
\end{figure}

\begin{figure*}[t]
\includegraphics[width=0.32\textwidth]{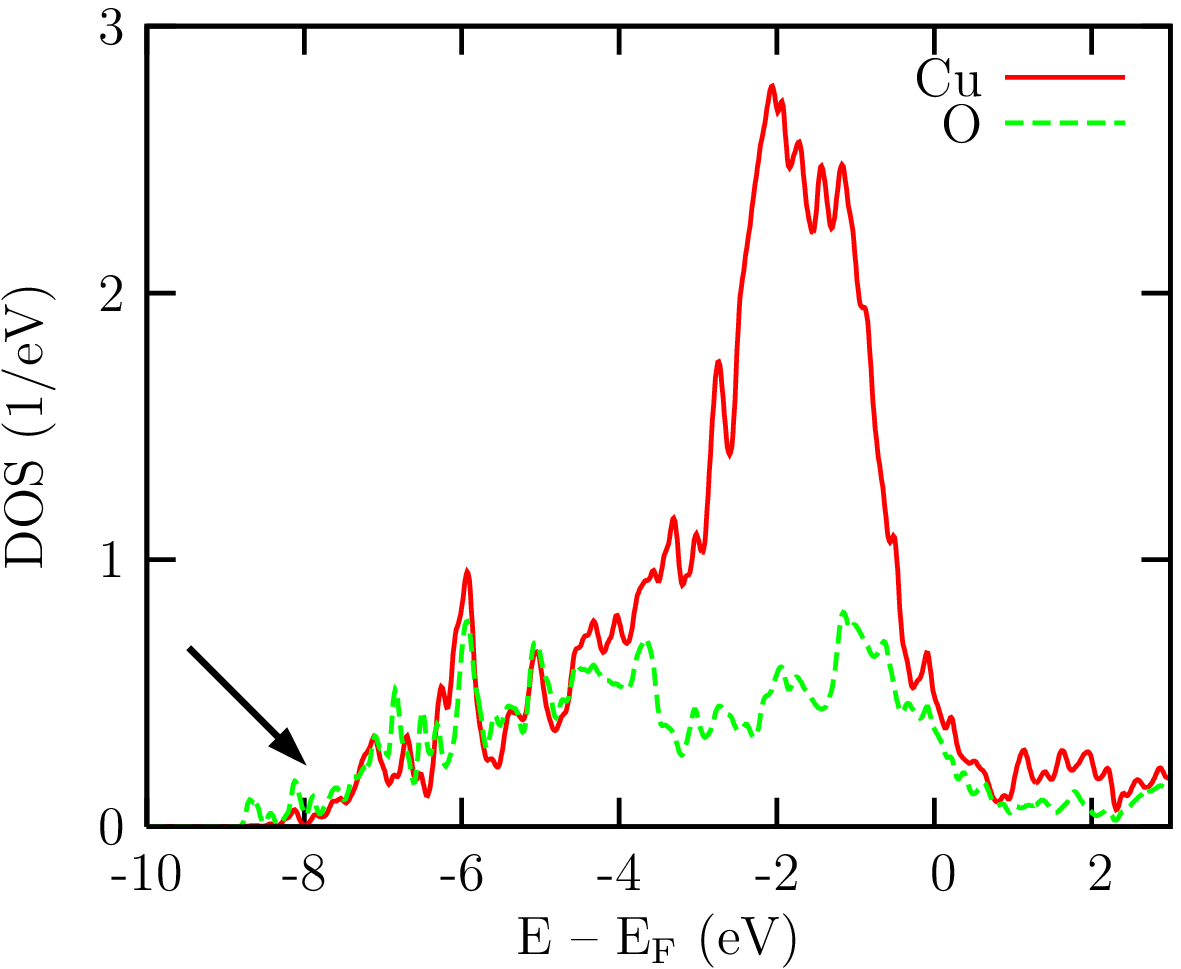}\hspace{0.2cm}
\includegraphics[width=0.32\textwidth]{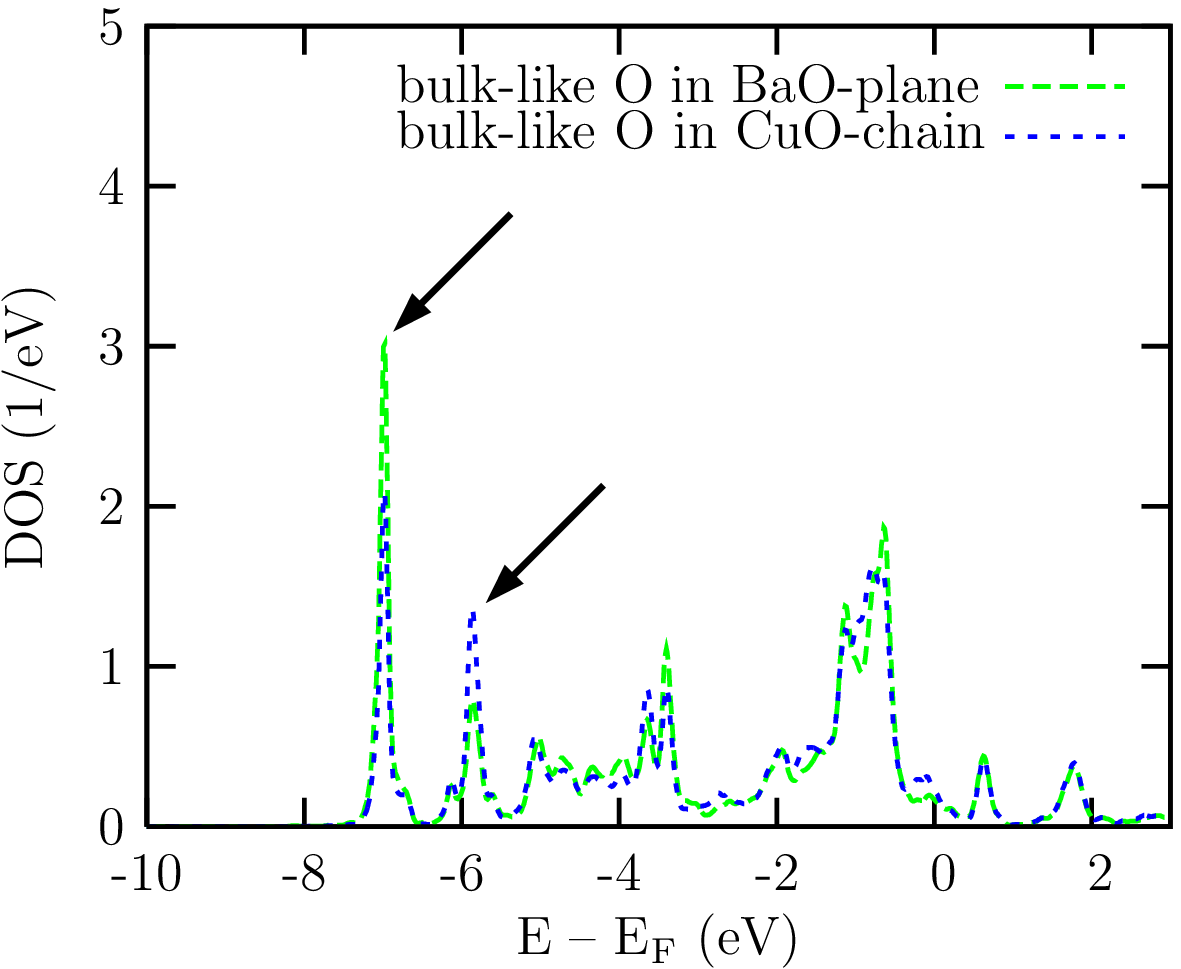}\hspace{0.2cm}
\includegraphics[width=0.32\textwidth]{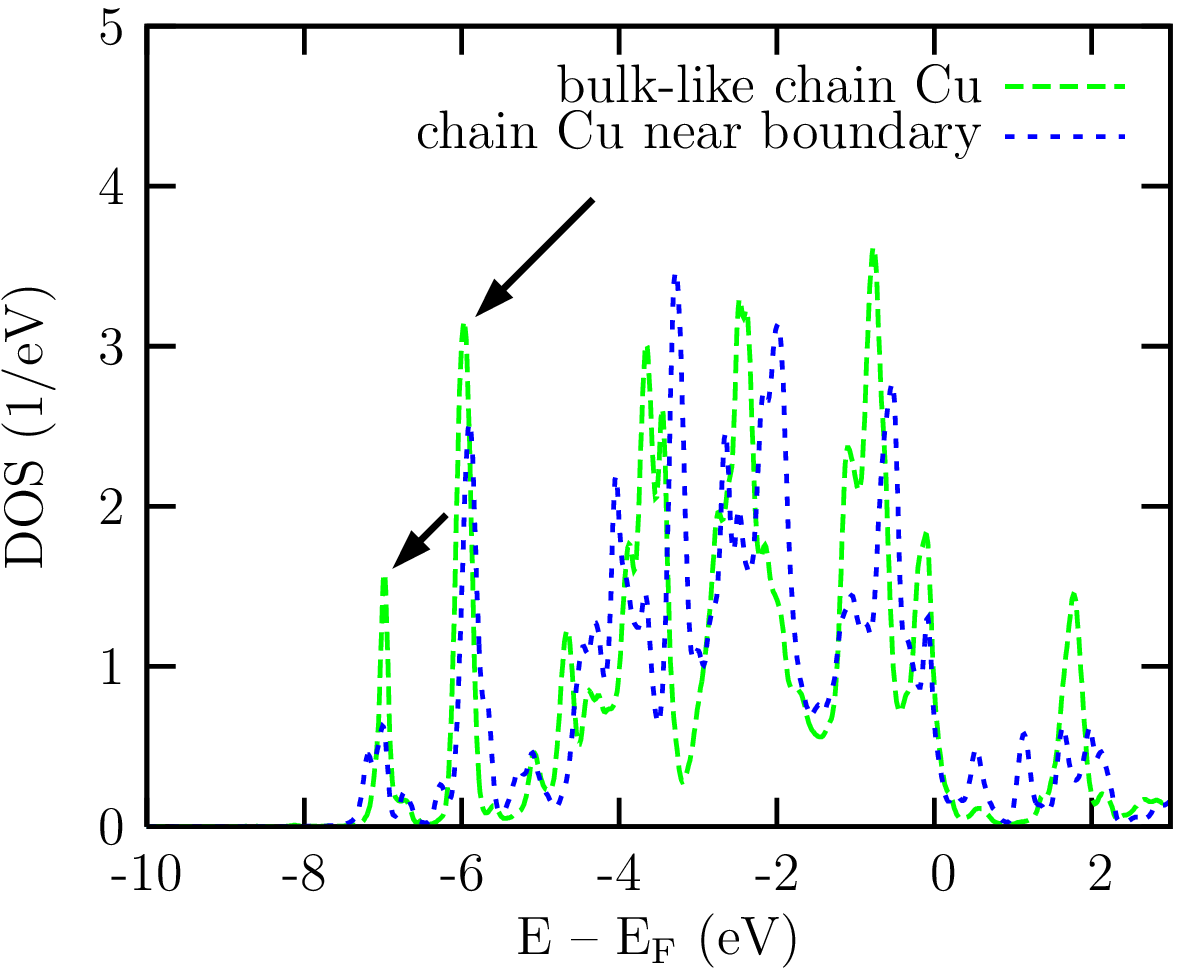}
\caption{(Color online) Left: partial Cu $3d$ and O $2p$ DOS for the entire supercell, normalized
by the number of Cu/O atoms. Center/right: site-projected DOS of selected
sites in the 45$^{\circ}$ YBCO grain boundary: bulk-like coordinated O
(center; showing two examples) and Cu at the grain boundary compared to bulk-like
coordinated Cu (right; both on CuO chain sites).}
\label{fig3}
\end{figure*}
Contrary to conventional superconductors, bending of the band structure by a variation
in the charge distribution is strong enough to control the transport in high-T$_c$
compounds \cite{hil02} because of large dielectric constants and small carrier densities
\cite{samara90}. The Thomas-Fermi screening length (over which a
band-bending is effective) therefore grows to the order of magnitude of the superconducting
coherence length. Consequently, the technical optimization of grain boundaries calls
for further insight into the local electronic structure. This issue is addressed in the present
letter for a 45$^\circ$-tilted [001] grain boundary. Since the electronic states
are expected to depend strongly on the local atomic environment, we use the DFT approach
to obtain and study a fully relaxed YBCO grain boundary in a supercell of 
altogether 317 atoms. Our data provide evidence for a characteristic charge transfer between the
CuO$_2$ chains and CuO chains. We explain this effect in terms of the modifications
of the chemical bonding.


Our results are based on density functional theory and the generalized gradient
approximation, as implemented in the WIEN$2k$ package \cite{wien2k}. By its
all-electron scheme, this full-potential linearized augmented-plane-wave code is suitable
for dealing with structural relaxation and the induced charge redistribution in complex
geometries, applying a supercell approach \cite{epl}. From a technical point of view,
the electronic states at YBCO grain boundaries are accessible to the supercell approach,
based on three-dimensional periodic boundary conditions, since relevant effects are proposed
to take place on the length scale of the lattice constant. Typical experimental screening lengths
for doped YBCO systems are $\sim 4.5$ \AA\ \cite{mannhart92}, which is an upper limit for the system under
investigation. A supercell therefore can cover the essentials of the electronic interaction,
if density functional theory predicts the correct screening length.

The supercell which we use for simulating a 45$^{\circ}$-tilted [001] grain
boundary is displayed in Fig.\ \ref{fig1}, in a projection parallel to the $c$-axis
of the parent YBCO unit cell (black square). Our structural setup reduces the number
of atomic sites entering the calculation to the necessary minimum. A further reduction
is not possible since one would not maintain both Cu and O atoms in a bulk-like
coordination. We mention that a (minor) lattice strain is present in our supercell by
construction. It results from a lattice mismatch of less than 1\% of the YBCO
lattice constants between the neighbouring grains in our supercell: five 45$^{\circ}$-tilted
and seven non-tilted YBCO unit cells form the grain boundary, compare Fig.\ \ref{fig1}.
The strain has to be accepted in order to keep periodic boundary conditions.
However, the mismatch is small enough to ensure that there is no drawback on the validity
of our results. For sampling the Brillouin zone a $5\times5\times2$ mesh with 25
points in the irreducible wedge is applied. We use 2502 local orbitals and $\approx840000$
plane waves to represent the charge density. The cutoff is set to RK$_{\rm max}=4$.


Turning to the results of our electronic structure calculations for the YBCO grain boundary,
we give in Fig.\ \ref{fig3} an overview of the valence states, consisting  of the Cu $3d$
and O $2p$ orbitals. The DOS curves which we present here and in the following are normalized
by the number of contributing atoms in order to simplify a comparison. The Cu $3d$
DOS exhibits a broad structure with one pronounced maximum centered at about $-2$ eV.
This shape reflects the YBCO bulk data except for the fact that the width of the $-2$ eV
peak is increased, which is expected since the different Cu atoms at a disordered
grain boundary should have slightly different chemical surroundings. Likewise similar
to bulk YBCO, there are substantial O $2p$ contributions in the entire Cu $3d$
energy range up to the Fermi level due to a strong Cu-O hybridization. The latter also
explains the Cu $3d$ admixtures at low energy. However, at this point we
encounter an important difference to bulk YBCO \cite{krakauer88}: A distinct
portion of electronic states appears below $-7$ eV. These states, of course, are related
to the modified bonding at the grain boundary.

\begin{figure*}[t]
\includegraphics[width=0.32\textwidth]{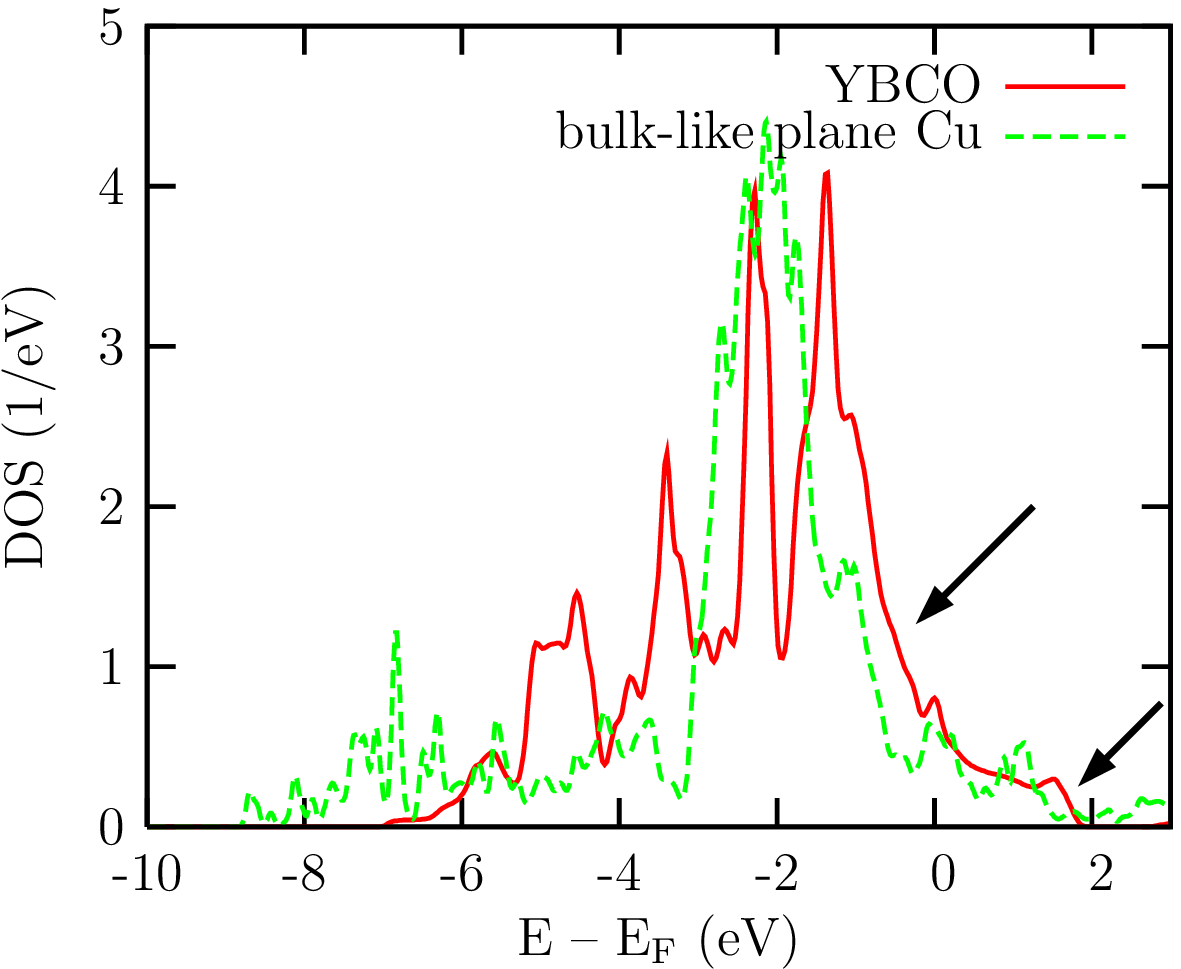}\hspace{0.5cm}
\includegraphics[width=0.32\textwidth]{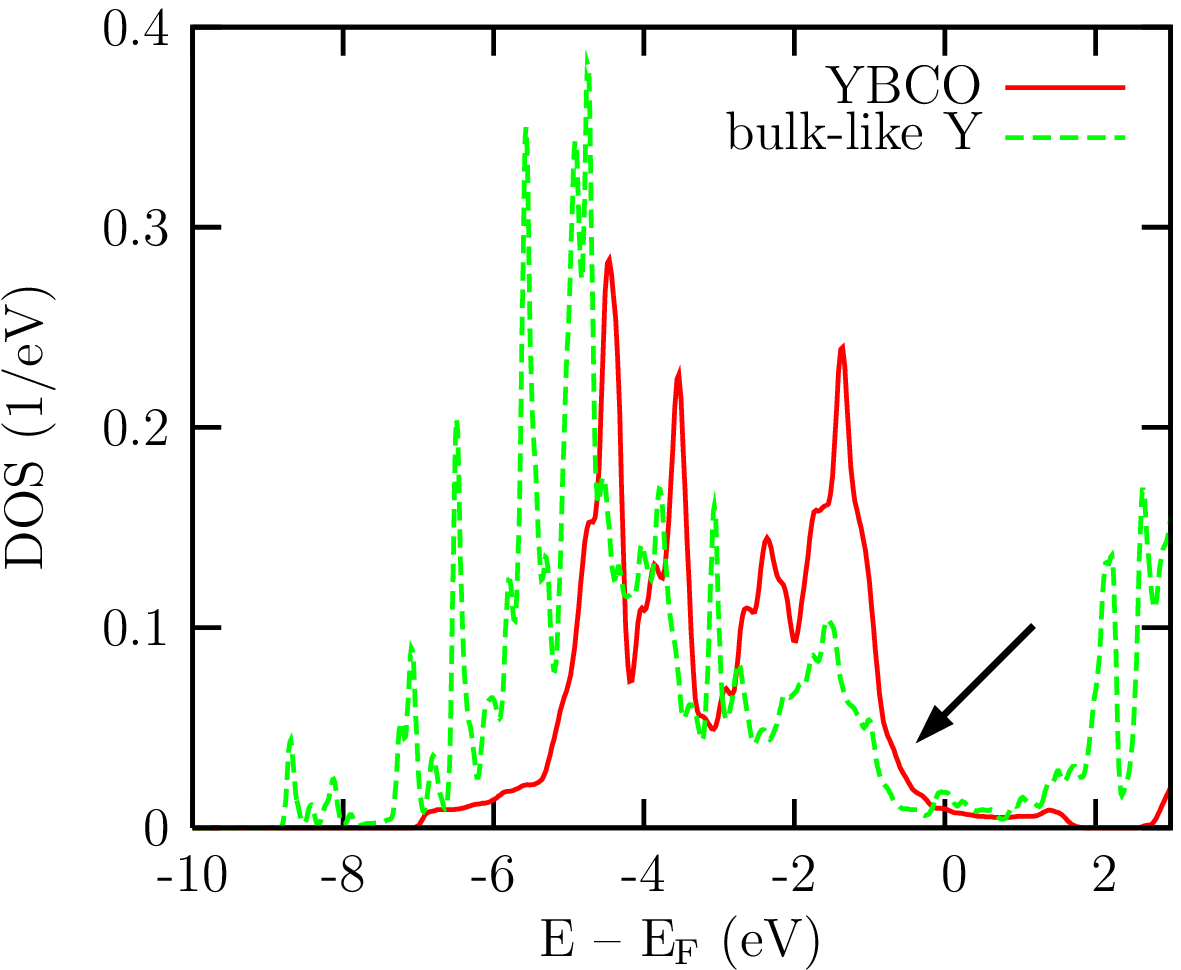}\\[0.3cm]
\includegraphics[width=0.32\textwidth]{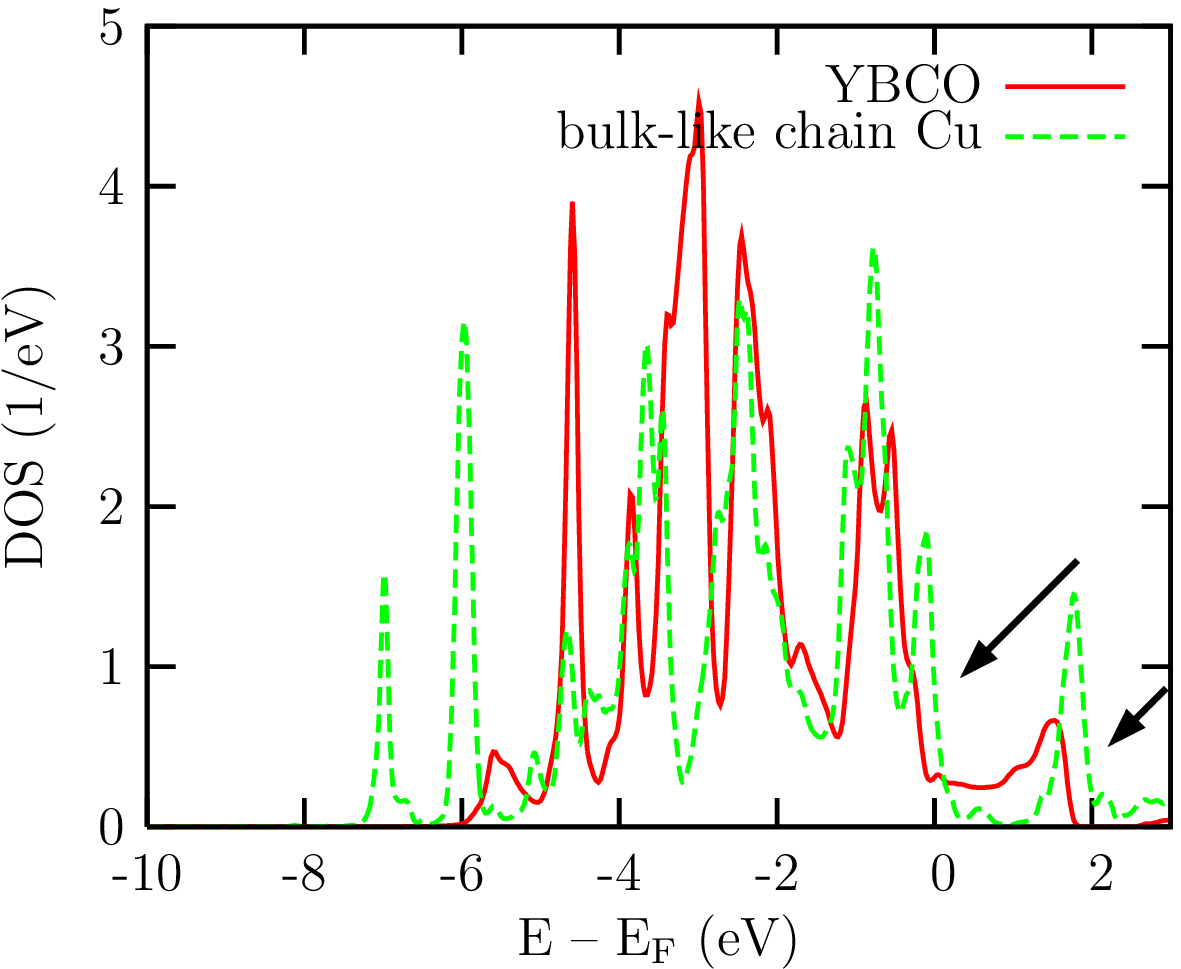}\hspace{0.5cm}
\includegraphics[width=0.32\textwidth]{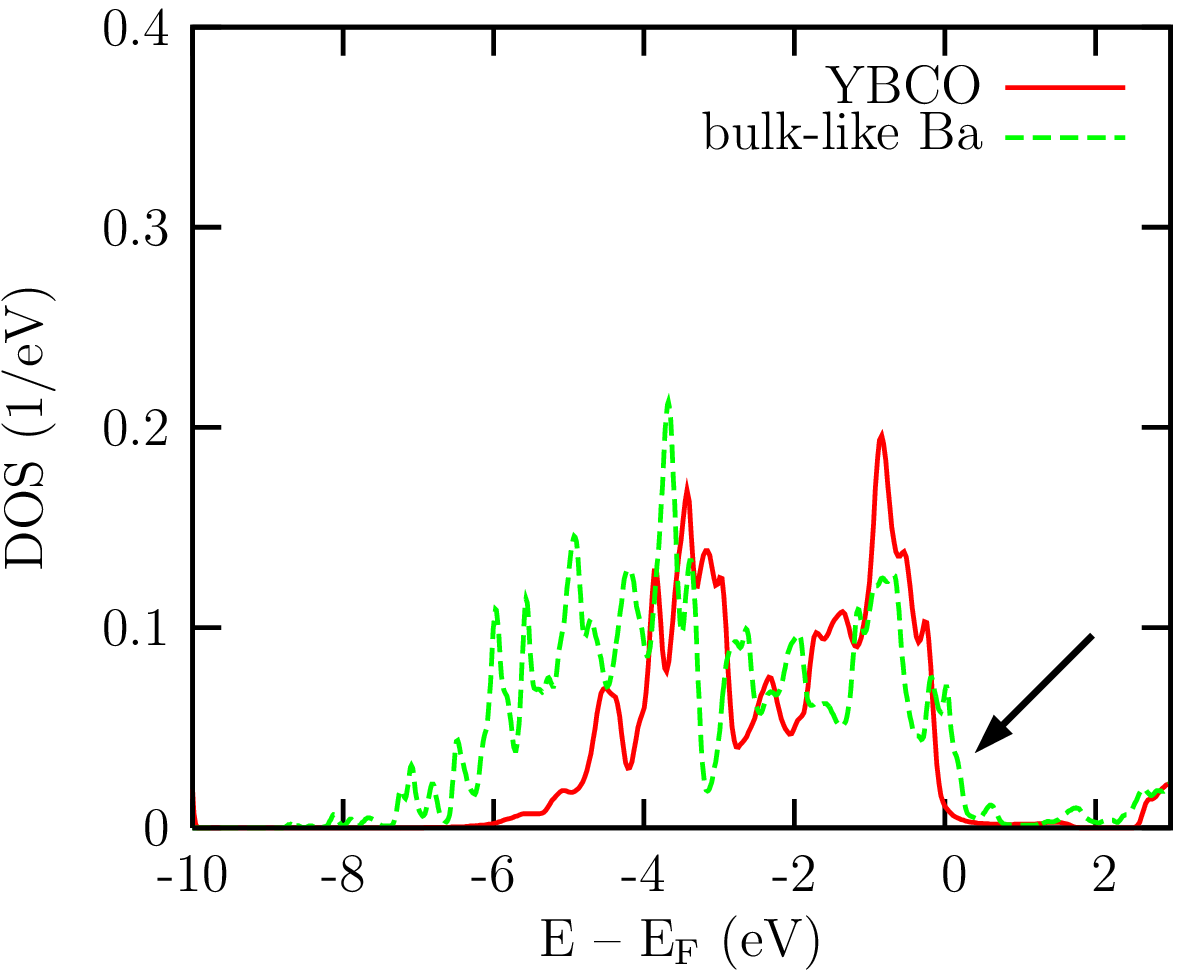}
\caption{(Color online) Partial (site-projected) DOS of selected bulk-like coordinated
atoms in the 45$^{\circ}$ YBCO grain boundary, as compared to the corresponding
YBCO bulk DOS: Cu in the CuO$_2$ planes (left top), Cu in the CuO chains (left bottom),
Y (right top), and Ba (right bottom).}
\label{fig4}
\end{figure*}

In order to analyze the Cu-O bonding in more detail, the DOS can be projected on
the atomic sites of interest. Corresponding results are given in Fig.\ \ref{fig3}.
In the central panel, we compare the
O $2p$ DOS of two (representative) bulk-like O atoms, one from a BaO layer and one from
a CuO chain. The term ``bulk-like'' is used to indicate that an atom is coordinated as in
bulk YBCO, i.e.\ its first coordination sphere is not affected
by the grain boundary. For all bulk-like O atoms we obtain a similar behaviour: Above
some $-5$ eV the DOS resembles the bulk YBCO DOS, whereas two additional peaks
appear at lower energy. These states trace back to additional Cu-O bonds created
at the grain boundary due to the broken translational symmetry, see Fig.\ \ref{fig1}.
Yet, because they reappear for bulk-like atoms they cannot be described in terms
of localized states at the grain-grain interface. A possible explanation would be a
substantial delocalization of the hybridized Cu-O states, which could reach deep into
the grain. However, our minimal supercell does not allow us to resolve the latter. In contrast to the naive expectation for $d$ electron orbitals,
such a delocalization seems to be typical for metallic YBCO \cite{temmerman01}. This
line of reasoning is supported by the Cu $3d$ DOS. On the right hand side of
Fig.\ \ref{fig3} we show, as a representative example, data for Cu atoms from a CuO
chain. Both for the bulk-like and the boundary atom the new low energy states are present.

It is well-established that the electronic characteristics of copper oxides depend severely
on details of the chemical bonding and on the doping \cite{filippetti05,US_prl}. Here,
our results show an interdependence between the modified chemical bonding
at the grain boundary and the charge distribution between the structural building blocks,
in particular the CuO$_2$ planes and CuO chains. While the creation of additional
CuO bonds predominantly affects the low energy range of the valence states, see the
previous discussion, Fig.\ \ref{fig4} demonstrates that there are likewise distinct alterations
at the Fermi energy ($E_F$). For selected bulk-like atoms in our supercell, i.e.\ Cu in the
CuO$_2$ planes (left top) and in the CuO chains (left bottom), Y (right top), and Ba
(right bottom), we study the site-projected DOS for the grain boundary (dashed line) in
comparison to the respective bulk YBCO results (solid line). Around $E_F$ we find
significant shifts of the energetic positions of the band edges, indicated by arrows
in Fig.\ \ref{fig4}.

\begin{figure}[t]
\includegraphics[width=0.165\textwidth,angle=-45]{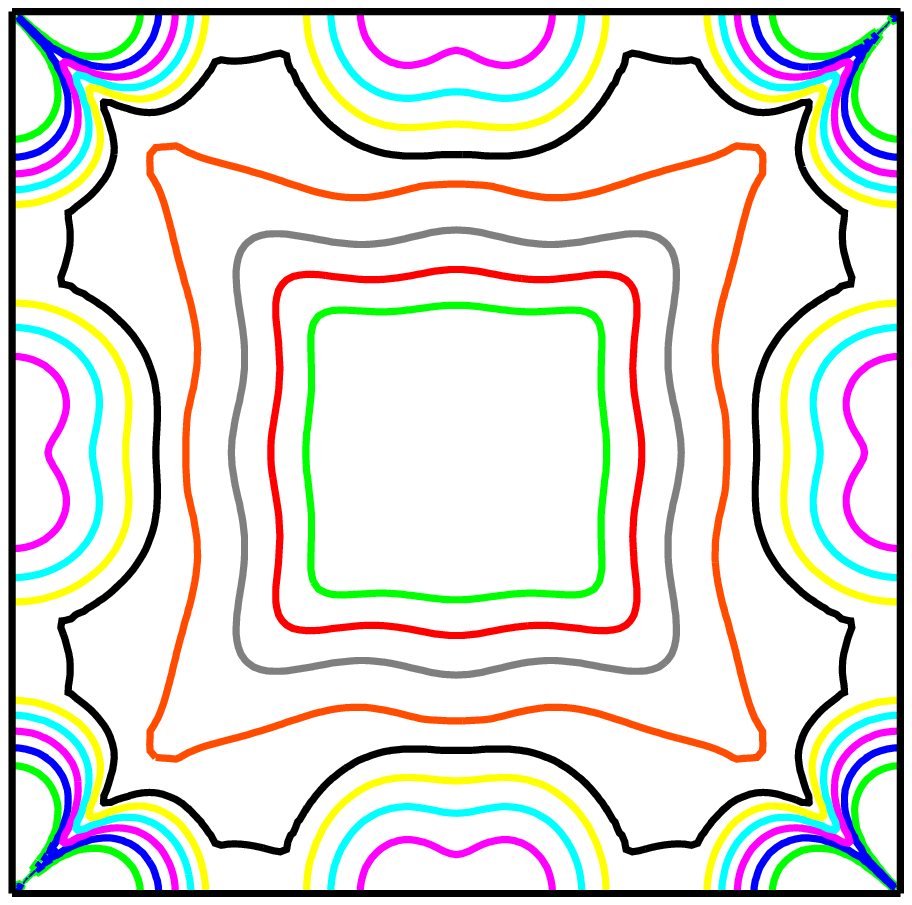}
\includegraphics[width=0.165\textwidth,angle=-45]{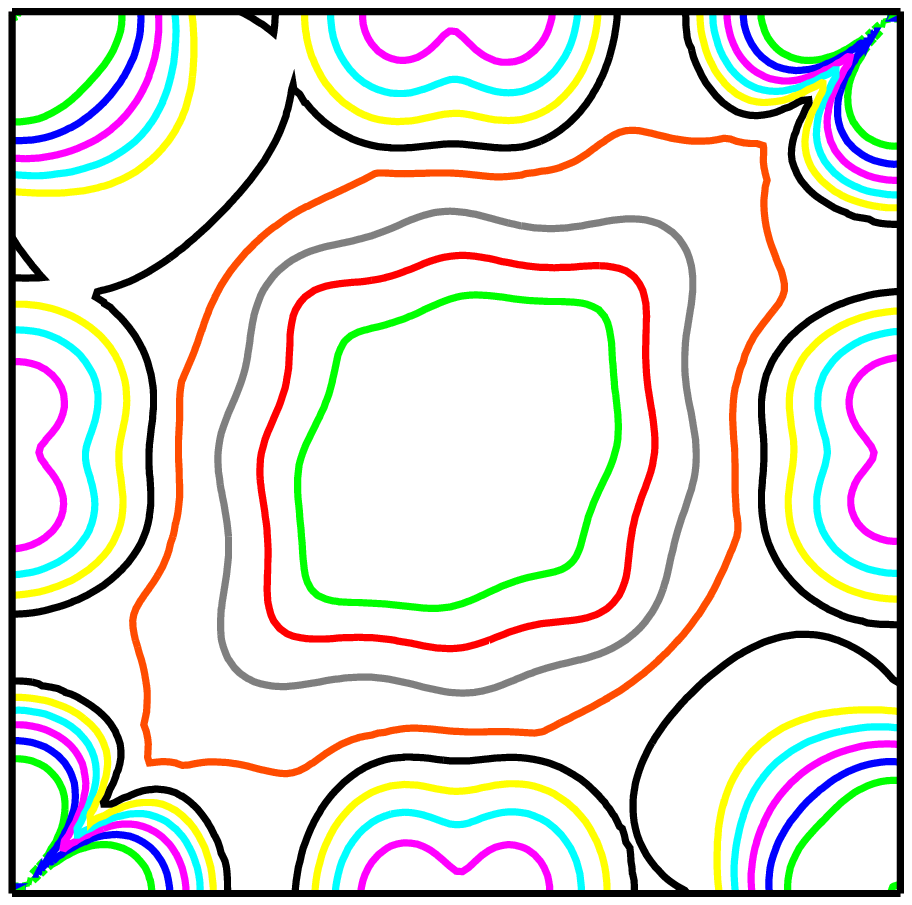}
\caption{(Color online) Contour plots of the hole density, i.e.\ the
integrated total DOS between the Fermi level and 1.9 eV, in the CuO$_2$ planes of
bulk YBCO (left) and of the supercell (right). The display area is the square shown
in Fig.\ \ref{fig1}, with Cu in the corners and O in the centers of the edges.}
\label{fig5}
\end{figure}

For the CuO$_2$ planes, the electronic states shift to the low energy side. The electron
count hence increases near the grain boundary, i.e.\ the hole count is reduced.
Quantitative evaluation yields a reduction of $0.10\pm0.02$ holes per Cu atom, which
is a substantial modification of the local doping state and should be accompanied by
strong changes in the electronic properties of the CuO$_2$ planes. With reference to
the YBCO phase diagram \cite{pickett89}, it is to be expected that this hole underdoping
will prohibit the transition of the grain boundary into a superconducting state. The
shift of the electronic states can also be interpreted as a compactification
of the Cu $3d$ DOS for sites at a grain boundary, see the left top panel of
Fig.\ \ref{fig4}. This corresponds to a reduced delocalization of the hybridized Cu-O
states, tracing back to the broken lattice periodicity. Hence, we attribute the charge
transfer towards the CuO$_2$ planes to an enhanced spatial localization of their near-$E_F$
states. Finally, the additional charge originates from the CuO chains, see the left bottom
panel of Fig.\ \ref{fig4}, since here the band edges shift in the opposite direction,
i.e.\ to higher energy. Due to the small portion of Y and Ba contributions to the valence
states, these sites play a minor role for the redistribution. The hole density
in the CuO$_2$ planes is shown in Fig.\ \ref{fig5} by contour plots for the area
of the square in Fig.\ \ref{fig1}. The observed hole reduction affects almost
the entire CuO$_2$ plane to a similar degree, where qualitative deviations of the
contours appear only perpendicular to the grain boundary.

In conclusion, we have studied a 45$^{\circ}$-tilted [001] grain
boundary in YBCO by density functional theory and performed a full structure optimization.
As to be expected, we obtain significant modifications of the Cu-O chemical bonding due
to additional combinations of effective overlap between Cu $3d$ and O $2p$ orbitals at
the grain-grain interface. Beyond, the bulk-like Cu-O states are subject to enhanced
localization near the grain boundary as the crystal lattice is perturbed. This
results in an energetic shift of the electronic states in the vicinity of the Fermi
level, which has almost the same amplitude in the CuO$_2$ planes and the CuO chains,
but the opposite sign.

As a consequence, there is a substantial charge transfer between these structural
building blocks. The CuO$_2$ planes are affected by a charge carrier (hole) depletion
of $0.10\pm0.02$ holes per Cu atom. A grain boundary, hence, can be interpreted in
terms of (intrinsically) underdoped CuO$_2$ planes, which explains a suppressed transition
into a superconducting state. In addition, it explains the fact that the supercurrent
density can be enhanced by a local hole overdoping of the grain boundaries
\cite{schmehl99,hammerl00}, since this compensates the intrinsic underdoping. Assuming
that a third of the additional holes enters the CuO$_2$ planes, the maximum of the
enhancement for 30\% Ca doping agrees even quantitatively with the obtained
first-principles results. Finally, going beyond experimental data
\cite{browning93,babcock94}, the charge redistribution is found to be {\it inhomogeneous}
with depletion and accumulation zones at the grain boundary, while there is little
charge transfer to/off the grain.
New experiments are indicated to confirm this observation.

\subsection*{Acknowledgement}
We acknowledge fruitful discussions with J.\ Mannhart
and financial support by the Deutsche Forschungsgemeinschaft (SFB 484).

\end{document}